\documentclass{appolb}

\usepackage{amssymb}
\usepackage{amsmath}
\usepackage{epsfig}
\usepackage{amsfonts}
\usepackage{graphicx}%

\begin{document}
\title{Thermodynamics of the $O(4)$ linear and nonlinear models within the auxiliary
field method%
\thanks{Presented at the Workshop \textquotedblleft Excited QCD 2011\textquotedblright\ 
, 20-25 February 2011-Les Houches }%
}
\author{Elina Seel
\address{Institute for Theoretical Physics, Johann Wolfgang Goethe University,
Max-von-Laue-Str.\ 1, D--60438 Frankfurt am Main, Germany}
}
\maketitle
\begin{abstract}
The study of the $O(N)$ model at nonzero temperature is presented applying the 
auxiliary field method, which allows to obtain a continuous transformation between
 the linear and the nonlinear version of the model. In case of explicitly broken 
chiral symmetry the order of the chiral phase transition changes from crossover to 
first order as the vacuum mass of the $\sigma$ particle increases. In the chiral limit
 one observes a first order phase transition and the Goldstone's theorem is fulfilled.
\end{abstract}
\PACS{PACS numbers: 11.10.Wx, 11.30.Rd, 12.39.Fe.}
  
\section{Introduction}
Scalar models with orthogonal symmetry are applied in many areas of physics,
like quantum dots and high-temperature superconductivity. In three spatial
dimensions no analytical solution exists, therefore it is instructive to
compare different many-body approximation schemes to estimate their physical
relevance. In the literature the optimized perturbation theory \cite{chiku},
the 2PI formalism \cite{dirk, petropoulos}, and the $1/N$ expansion
\cite{coleman, meyer, bochkarev, brauner} have been used several times to
examine the thermodynamical behavior of the $O(N)$ model.

In this work we study the thermodynamics of the $O(N)$ model by introducing an
auxiliary field. To calculate the effective potential, the masses and the
condensate at nonzero $T$ we apply the so-called two-particle irreducible
(2PI) or Cornwall-Jackiw-Tomboulis (CJT) formalism \cite{cornwall, kadanoff}
in the double-bubble approximation. Within the auxiliary field method the
nonlinear version of the model is given by a mathematically well-defined
limiting process of the linear $O(N)$ model. We find that the gap equations
for the order parameter and the masses of $\sigma$ and $\pi$ quantitatively
differ from the standard treatment of the $O(N)$ model without the auxiliary
field. This paper is based on the results of Ref. \cite{lvnl}.

\section{The $O(N)$ model}

The generating functional at finite temperature of the linear $O(N)$ model is
given by%
\begin{equation}
\ Z_{L}[\varepsilon,h]=N\int\mathcal{D}\alpha\mathcal{D}\Phi e^{\overset
{\ }{\int_{0}^{\beta}}d\tau\int_{V}^{\ }d^{3}x\mathcal{L}_{\sigma
\text{-}\alpha}}\text{ },\label{gf1}%
\end{equation}%
\[%
\begin{array}
[c]{ccc}%
\mathcal{L}_{\sigma\text{-}\alpha}=\dfrac{1}{2}\partial_{\mu}\Phi^{t}%
\partial^{\mu}\Phi-U(\Phi,\alpha) &  & \text{ }U(\Phi,\alpha)=\dfrac{i}%
{2}\alpha(\Phi^{2}-\upsilon_{0}^{2})+\dfrac{\varepsilon}{2}\alpha^{2}%
-h\sigma\text{ .}%
\end{array}
\]
Here $\Phi^{t}=\left(  \sigma,\pi_{1}...\pi_{N}\right)  $; $\alpha$ is an
auxiliary field serving as a Lagrange multiplier. By integrating it out 
we obtain:%
\begin{equation}
\ Z_{L}[\varepsilon,h]=\int\mathcal{D}\Phi e^{\overset{\ }{\int_{0}^{\beta}%
}d\tau\int_{V}^{\ }d^{3}x\left[  \frac{1}{2}\partial_{\mu}\Phi^{t}%
\partial^{\mu}\Phi-\frac{1}{2N\varepsilon}\left(  \Phi^{2}-\upsilon_{0}%
^{2}\right)  ^{2}+h\sigma\right]  }\text{ }.\label{aux}%
\end{equation}
The tree level potential exhibits now the typical \textquotedblleft Mexican
hat\textquotedblright\ shape, where $1/\varepsilon$ is the coupling constant,
$h$ the parameter for explicit symmetry breaking, and $\upsilon_{0}$ is the
vacuum expectation value (v.e.v.). The advantage of the auxiliary field
representation of the $linear$ version of the model, Eq. (\ref{aux}), is that
by taking the limit $\varepsilon\rightarrow0$ one naturally recovers the
$nonlinear$ version of the model. Note, the limit $\varepsilon\rightarrow0$
corresponds to an infinitely large coupling constant. In the nonlinear case
the dynamics of the fields is constrained on the chiral circle, defined by the
condition $\Phi^{2}=\upsilon_{0}^{2}$, which is represented by a $\delta
$-function
\begin{equation}
\ Z_{NL}[h]=\lim_{\varepsilon\rightarrow\text{ }0^{+}}Z_{L}[\varepsilon
,h]=\int\mathcal{D}\Phi\delta(\Phi^{2}-\upsilon_{0}^{2})e^{\overset{\ }%
{\int_{0}^{\beta}}d\tau\int_{V}^{\ }d^{3}x\left(  \dfrac{1}{2}\partial_{\mu
}\Phi^{t}\partial^{\mu}\Phi+h\sigma\right)  }\text{ .}%
\end{equation}
Here we have used the mathematically well-defined (i.e., convergent)
representation of the functional $\delta$-function%
\begin{equation}
\delta(\Phi^{2}-\upsilon_{0}^{2})=\lim_{\varepsilon\rightarrow\text{ }0^{+}%
}N\int\mathcal{D}\alpha\,e^{-\int_{0}^{1/T}d\tau\int_{V}d^{3}x\,\left[
\dfrac{i}{2}\alpha(\Phi^{2}-\upsilon_{0}^{2})+\dfrac{N\varepsilon}{8}%
\alpha^{2}\right]  }\text{ . }\label{delta}%
\end{equation}
In some previous studies of the nonlinear $O(N)$ model \cite{meyer, bochkarev}
the $\varepsilon$-dependence of the $\delta$-function was not properly
handled, since the $\varepsilon$-dependent term, $\varepsilon\,\alpha^{2},$
was neglected. This is, however, not correct, since this term is essential to
construct the link between the linear and the nonlinear versions of the model.
Besides, an integration over the auxiliary field does not give the correct
potential of the linear model when this term is absent.

\section{The effective potential and gap equations}

The effective potential within the CJT formalism is given by
\begin{equation}
V=U(\phi)+\dfrac{1}{2}\int_{k}[\ln G^{-1}(k)+D^{-1}(k;\phi)G(k)-1]\text{
}+V_{2}(\phi,G)\text{ .}%
\end{equation}
Here $U(\phi)$ is the tree-level potential, $D(k;\phi)$ the tree-level
propagator in momentum space, $G(k)$ the full propagator in momentum space,
and $V_{2}(\phi,G)$ contains all two-particle irreducible diagrams.\textbf{
}In our case the tree-level potential is given by%
\begin{equation}
U=-\dfrac{i}{2}(\alpha_{0}+\alpha)(\sigma^{2}+\pi_{i}^{2}+2\sigma\phi
\ +\phi^{2}-\upsilon_{0}^{2})-\dfrac{N\varepsilon}{8}(\alpha_{0}+\alpha
)^{2}+h(\phi+\sigma)\text{ },
\end{equation}
where the fields $\sigma$\ and $\alpha$\ have been shifted around their
non-vanishing vacuum expectation values, $\sigma\rightarrow\phi+\sigma$ and
$\alpha\rightarrow\alpha_{0}+\alpha$. These shifts generate a bilinear mixing
term, $i\alpha\sigma\phi,$ rendering the mass matrix non-diagonal in the
fields $\sigma$\ and $\alpha$. Performing a further shift of the auxiliary
field $\alpha,\ \alpha\rightarrow\alpha-4i\phi\sigma/N\varepsilon$ $,$ this
unphysical mixing can be eliminated. The resulting Lagrangian contains no
4-point vertices%
\begin{align}
\mathcal{L}_{\sigma\text{-}\alpha}  &  =\dfrac{1}{2}\partial_{\mu}%
\sigma\partial^{\mu}\sigma+\dfrac{1}{2}\partial_{\mu}\pi_{i}\partial^{\mu}%
\pi_{i}-\dfrac{\sigma^{2}}{2}\left(  i\alpha_{0}+4\dfrac{\ \phi^{2}%
}{N\varepsilon}\right)  -\dfrac{\pi_{i}^{2}}{2}\left(  i\alpha_{0}\right)
\nonumber\\
&  -\dfrac{1}{2}\,\frac{N\varepsilon}{4}\,\alpha^{2}-\dfrac{i}{2}\alpha
(\sigma^{2}+\pi_{i}^{2})-\dfrac{2\ \phi}{N\varepsilon}\sigma(\sigma^{2}%
+\pi_{i}^{2})\nonumber\\
&  -\dfrac{i}{2}\alpha_{0}(\phi^{2}-\upsilon_{0}^{2})-\dfrac{N\varepsilon}%
{8}\alpha_{0}^{2}+h\phi\text{ }.
\end{align}
Therefore, if we restrict ourselves to the so-called double-bubble
approximation where the self-energy of the particles is independent of
momentum, the contribution of $V_{2}$\ to the CJT effective potential vanishes.

The gap equations are derived by minimizing the effective potential and read:
\begin{align}
h &  =\phi\left[  M_{\pi}^{2}(\varepsilon,h)+\dfrac{4}{N\varepsilon}\int
_{k}G_{\sigma}(k)\right]  \text{ },\text{ \ }M_{\sigma}^{2}\left(
\varepsilon,h\right)  =M_{\pi}^{2}\left(  \varepsilon,h\right)  +\dfrac
{4\phi^{2}}{N\varepsilon}\text{ },\nonumber\\
M_{\pi}^{2}\left(  \varepsilon,h\right)   &  =\frac{2}{N\varepsilon}\left[
\phi^{2}-\upsilon_{0}^{2}+\int_{k}G_{\sigma}(k)+(N-1)\int_{k}G_{\pi
}(k)\right]  \text{ }.\label{gap}%
\end{align}

\section{\textbf{Results}}

The numerical results are presented for $N=4$ corresponding to a system of
three pions and\ their chiral partner, the $\sigma$ particle. We apply the
trivial renormalisation (TR), where the divergent vacuum contributions of the
tad-pole diagrams is set to zero.

\begin{figure}
[ptb]
\begin{center}
\includegraphics[
height=1.9104in,
width=4.209in
]%
{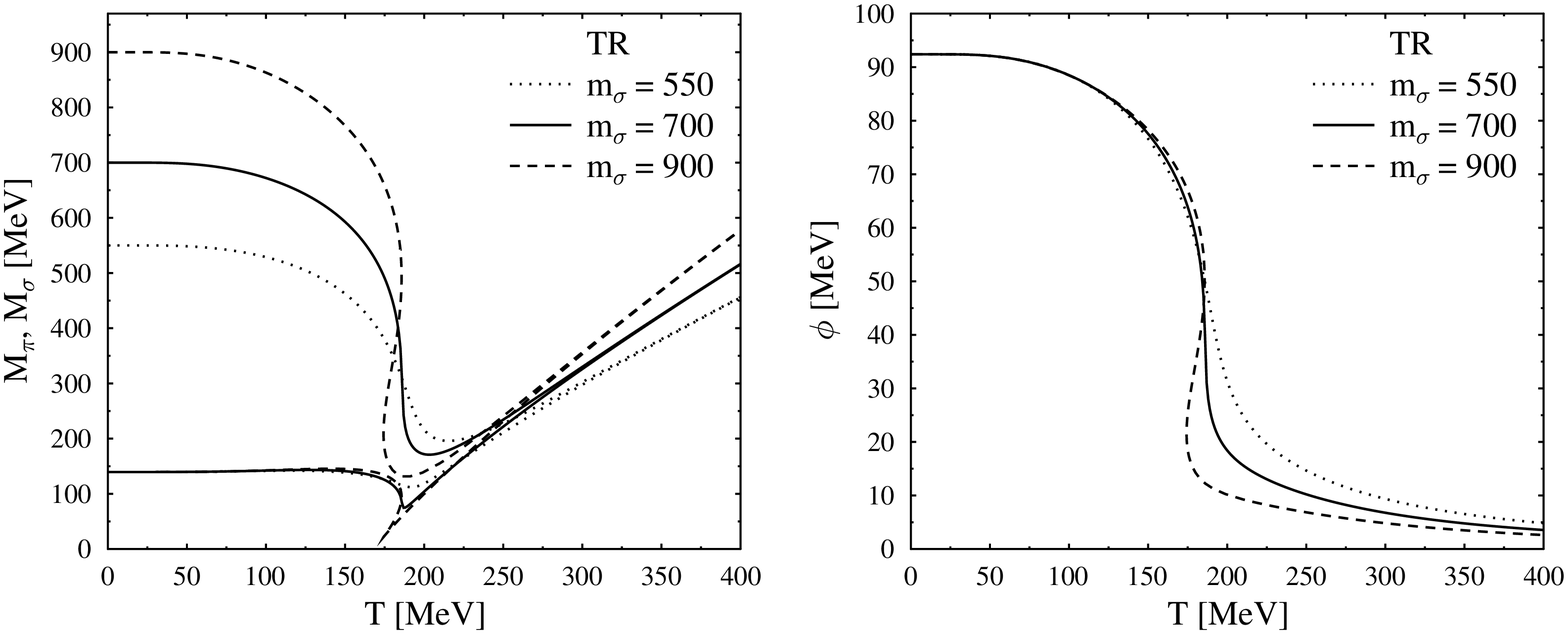}%
\caption{The pion mass, the sigma mass and the condensate as a function of $T$
in the $O(4)$ linear model in case of explicitly broken symmetry for different
values of $m_{\sigma}$.}%
\label{test}%
\end{center}
\end{figure}

In the linear version of the model and for explicitly broken chiral symmetry,
the order of the chiral phase transition depends sensitively on the vacuum
mass of the $\sigma$ particle, $m_{\sigma}$, see Fig. \ref{test}. Increasing
$m_{\sigma}$the phase transition changes from crossover to first order. The 
identification of the chiral partner of the pion is under debate,
e.g. Refs. \cite{amslerrev}.

\begin{figure}
[ptb]
\begin{center}
\includegraphics[
height=1.8853in,
width=4.5455in
]%
{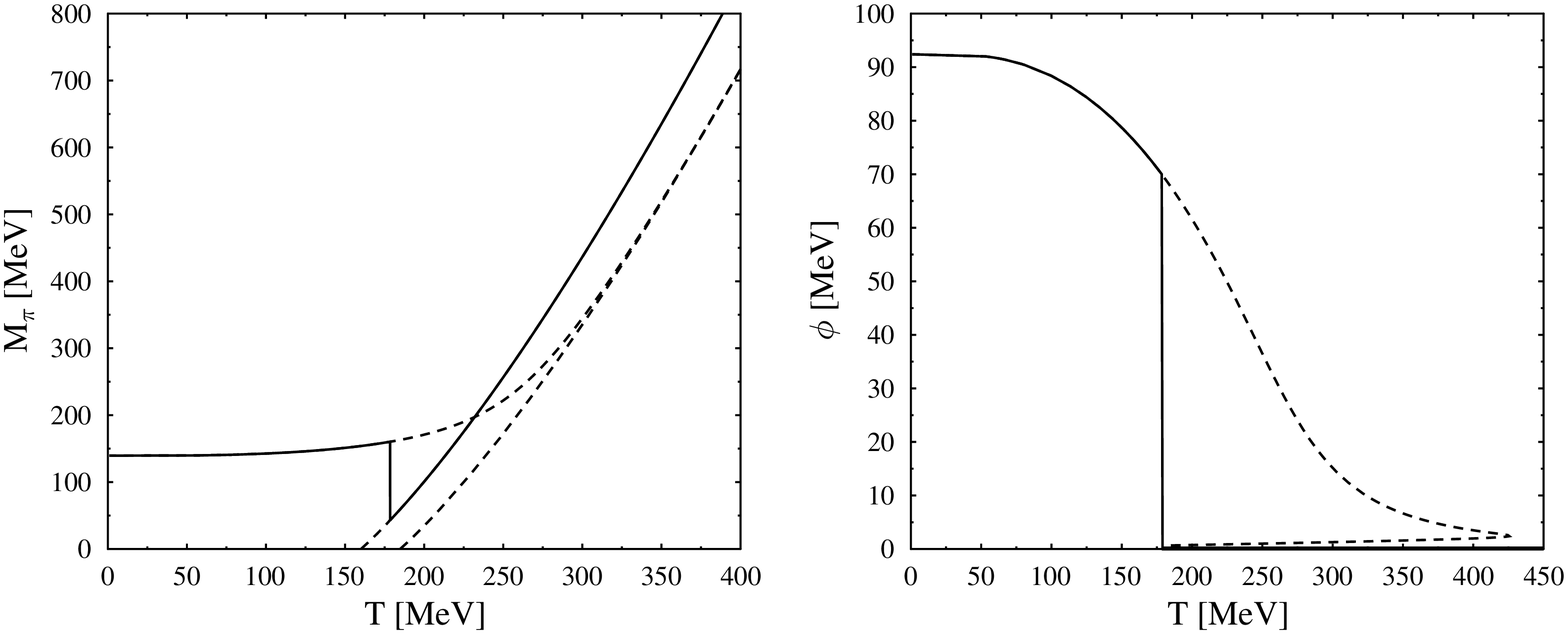}%
\caption{The pion mass and the condensate as a function of $T$ in the $O(4)$
nonlinear model in case of explicitly broken symmetry for $m_{\sigma
}\rightarrow\infty$ (in practice $m_{\sigma}=250$ MeV is used).}%
\label{h}%
\end{center}
\end{figure}

\begin{figure}
[ptb]
\begin{center}
\includegraphics[
height=1.8879in,
width=4.5446in
]%
{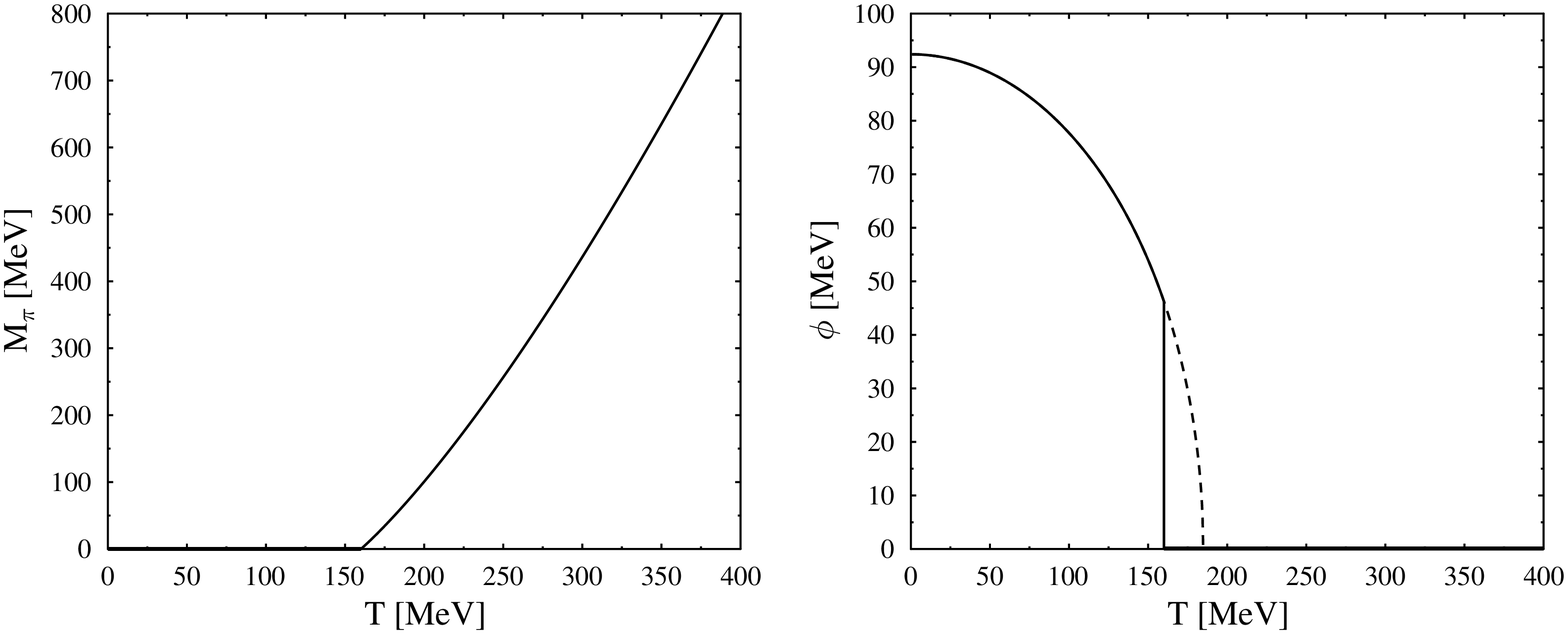}%
\caption{The pion mass and the condensate as a function of $T$ in the chiral
limit for $m_{\sigma}\rightarrow\infty$ (in practice $m_{\sigma}=250$ MeV is used).}%
\label{cl}%
\end{center}
\end{figure}

Performing the nonlinear limit, $\varepsilon\rightarrow0,$ one observes a
first order phase transition for explicitly broken chiral symmetry with the
critical temperature $T_{c}=178.6$ MeV, see\textbf{ }Fig.\ \ref{h}. In the
chiral limit, the phase transition is again of first order, see Fig.\ \ref{cl}%
, with $T_{c}=\sqrt{12/N}\,f_{\pi}$ $=\sqrt{3}\,f_{\pi}$, where $f_{\pi}=92.4$
MeV is the pion decay constant. In the phase where the symmetry is
spontaneously broken the pions are massless. Thus the Goldstone's theorem is
respected. Note that from the second equation in (\ref{gap}) the following
relation $1/\varepsilon=\left(  m_{\sigma}^{2}-m_{\pi}^{2}\right)  /\phi^{2}$
can be obtained. Thus, the nonlinear limit is equivalent to sending
$m_{\sigma}$ to infinity$.$

\section{Conclusions}

The study of the $O(N)$ model at nonzero $T$ was presented using the auxiliary
field method to construct a mathematically well defined link between the
linear and nonlinear versions of the model. To derive the thermodynamic
quantities like the effective potential, the temperature dependent masses and
the condensate we applied the CJT formalism in the double-bubble
approximation. Although qualitatively similar to the standard double-bubble
approximation in the treatment without auxiliary field, the gap equations are
quantitatively different and lead to different results for the order parameter
and the masses of the particles as a function of $T$.

A natural next step is to include sunset-type diagrams in the 2PI effective
action, which lead to nonzero imaginary parts for the self-energy of the
quasiparticles and, in turn, to a nonzero decay width. Another project is to
extend the studies to nonzero chemical potentials \cite{brauner} or to include
additional scalar states, since the nature of their constituency is quite
unclear \cite{achim}. Besides, the application of the auxiliary field method
should also be instructive for more complicated systems incorporating
additional vector and axial vector mesonic degrees of freedom \cite{stefan}
.\newline

\textbf{Acknowledgement}

The author thank S.\ Strueber, F.\ Giacosa, D. H.\ Rischke, T.\ Brauner, M.\ Grahl,
 A.\ Heinz, S.\ Leupold, and H.\ Warringa for interesting discussions. The work of 
E.\ Seel was supported by the Helmholtz Research School \textquotedblleft H-QM\textquotedblright.

%

\addcontentsline{toc}{section}{Abbildungs- und Literaturverzeichnis}%

\end{document}